\documentclass[twocolumn,aps,prl,showpacs,superscriptaddress,final]{revtex4-1}   

\usepackage{hyperref}
\hypersetup{
    colorlinks = true,
    linkcolor  = blue,
    citecolor  = blue,
    filecolor  = magenta,
    urlcolor   = cyan
}
\usepackage [english]{babel}
\usepackage [utf8]{inputenc}
\usepackage [T1]{fontenc}
\usepackage{graphicx}
\usepackage{amsmath}
\usepackage{amssymb}
\usepackage{siunitx}
\usepackage{booktabs}
\usepackage{float}
\usepackage{array}
\usepackage[normalem]{ulem}
\usepackage[dvipsnames]{xcolor}
\usepackage{dsfont}
\usepackage{siunitx}
\usepackage{soul}
\usepackage{dcolumn}
\usepackage{bm}
\usepackage{braket}
\usepackage{subfigure}

\newcommand{\kb}{\vec{k}}

\begin{document}
\setlength{\abovedisplayskip}{4pt}
\setlength{\belowdisplayskip}{4pt}
\setlength{\abovedisplayshortskip}{0pt}
\setlength{\belowdisplayshortskip}{0pt}
\preprint{APS/123-QED}

\title{Dipole-dipole interactions of Floquet states}

\author{Tim Ehret}
\author{Vyacheslav Shatokhin}
\author{Andreas Buchleitner}
\affiliation{
 Albert-Ludwigs-Universität Freiburg, Hermann-Herder-Str. 3, D-79104 Freiburg, Germany
}

\date{\today}

\begin{abstract}
We formulate a Floquet-Markov Lindblad master equation for translationally cold
two-level atoms driven by a strong monochromatic wave and coupled to
a common electromagnetic bath. The resulting
dipole-dipole interaction reproduces the anisotropic Heisenberg model. 

\end{abstract}

\maketitle
   
\emph{ Introduction.---}
Long-range dipole-dipole interactions 
are of central importance in physics and  chemistry, providing the microscopic mechanism for energy transport and collective phenomena in diverse many-body systems, from frozen Rydberg gases  \cite{Gallagher:98,Mourachko:1998}, over cold \cite{Bienaime:2012,Skipetrov:2014} and thermal atomic ensembles \cite{Bruder:2019,Ames:2022}, to light-harvesting complexes \cite{Scholes:2011,Scholak:2011}. Therefore, there has been steady interest in controlling the dipolar interactions, for instance, by changing the interparticle separations. With Rydberg atoms, a regime of dipole blockade can be achieved for sufficiently close atoms, making them a promising platform for quantum computation \cite{Jaksch:2000,Saffman:2010} and the simulation of many-body systems \cite{Browaeys_2020}. 

Another control strategy is to tune the atomic transition frequencies with an external periodic field \cite{Petrus_2008,Lee:2017}. The latter induces an AC Stark shift of the atomic energy levels \cite{Cohen_Tannoudji_atomphoton} and drives distinct dipole transitions into resonance, enabling the excitation transfer via the flip-flop process \cite{Petrus_2008,Lee:2017}, associated with multipartite entanglement resonances \cite{sauer_2012}. For Rydberg atoms, which exhibit a large polarizability \cite{Gallagher_1994} and ensure strong atom-laser interaction for microwave fields of moderate strength \cite{Bayfield_1974}, the shifts can be accurately assessed within the Floquet formalism \cite{Shirley:1965, Blümel_1987,Buchleitner_1995}. However, this strategy has so far not explored the potential change in the qualitative form of the dipole-dipole interactions in the presence of external driving. 

\begin{figure*}[t]
\includegraphics[width=.8\textwidth]{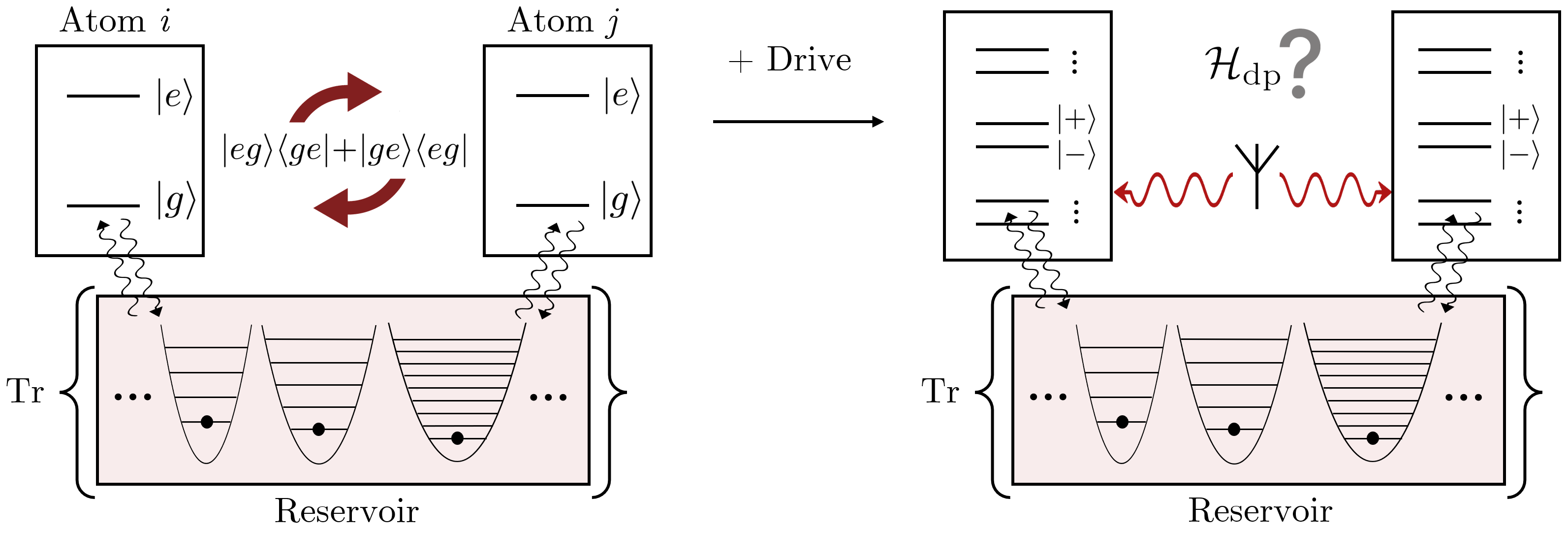}
\caption{Illustration of the physical setting. In the unperturbed case on the left-hand side, tracing over the degrees of freedom of the reservoir gives rise to an effective flip-flop interaction between the atoms, which transfers excitations. On the right-hand side, the presence of a strong drive (represented by the antenna) acting on the atoms modifies the atomic structure, inducing sidebands which offer new resonant transitions for the reservoir to couple. This leads to a modified dipole-dipole interaction Hamiltonian $\mathcal{H}_\mathrm{dp}$, given in Eq. \eqref{eq:6}.}
\label{fig:system}
\end{figure*}

We here address this problem, treating driven atoms as an open quantum system embedded into a common electromagnetic reservoir \cite{Lehmberg:1970}. We then formulate a Floquet-Markov Lindblad master equation \cite{Bluemel:1991} that indeed features a modified form of the dipole-dipole interactions. We attribute the new interaction terms to the emergence of sidebands in the (quasi)energy spectrum of the driven system, which enable additional couplings to the reservoir (see Fig.~\ref{fig:system}). Furthermore, we establish the equivalence of the new dipole-dipole Hamiltonian to that of the anisotropic Heisenberg model which has attracted much attention in the context of Floquet engineering of many-body spin systems \cite{Weitenberg:2021,Goldman:2015,Geier:2021,Scholl:2022,Nguyen_2024}. We therefore stress that our dipole-dipole interaction Hamiltonian is based on different physical principles than those used in Floquet engineering \cite{Weitenberg:2021,Goldman:2015,Geier:2021,Scholl:2022,Nguyen_2024}, where one maps the familiar flip-flop dipole-dipole interaction onto the target Hamiltonian with a sequence of time-delayed pulses, while considering the atoms as a closed quantum system.

\emph{Hamiltonian. ---}
We consider a system of $N$ identical, translationally cold two-level atoms at positions $\vec{r}_{i}$ with dipole moment $\vec{d}$ and transition frequency $\omega_{eg}$ between the ground and excited states, $\ket{g}$ and $\ket{e}$, respectively, that are driven by a classical monochromatic field, $\vec{E}_0\cos(\omega t)$, and coupled to the common radiation field at temperature $T_{\rm R}$. The total Hamiltonian is then  $\mathcal{H}=\mathcal{H}_\mathrm{S}+\mathcal{H}_\mathrm{R}+\mathcal{H}_\mathrm{I}$,
where (we measure the action in units of $\hbar $ throughout, i.e., $\hbar\equiv1$)  
\begin{align}
    \mathcal{H}_\mathrm{S}(t)&=\sum_{i=1}^N\left( \Omega_\mathrm{R}\cos(\omega t)\sigma^x_i + \omega_{eg}\sigma_i^z/2 \right),\label{eq:1}\\
    \mathcal{H}_\mathrm{R}&=\sum_{\Vec{k},s} \omega_{k}\left(a_{\vec{k}s}^\dagger a_{\vec{k}s}+1/2\right),\label{eq:2}\\
    \mathcal{H}_\mathrm{I}&=-i\sum_{i=1}^N\sum_{\vec{k},s} g_{\vec{k}s}^{i} \sigma^x_i\otimes \left(a_{\vec{k}s}-a^\dagger_{\vec{k}s}\right),
    \label{eq:3}
\end{align}
with $\sigma_i^x=\ket{e}_i\!\bra{g}_i+\ket{g}_i\!\bra{e}_i$ and $\sigma_i^z=\ket{e}_i\!\bra{e}_i-\ket{g}_i\!\bra{g}_i$ the atomic operators, $\Omega_\mathrm{R}=\vec{d}\vec{E}_0$ the Rabi frequency, $a_{\vec{k}s}$ ($a^\dagger_{\vec{k}s}$) the bosonic annihilation (creation) operators, and $g^i_{\vec{k}s}=\sqrt{\omega_k/(2 \epsilon_0  V)} (\vec{d}\vec{e}_{\kb s}) \exp({i\vec{k}\vec{r}_i})$ the coupling constants, with unit polarization vectors $\vec{e}_{\vec{k}s}\perp \vec{k}$ \cite{phases}. The Hamiltonian in Eqs. \eqref{eq:1}-\eqref{eq:3} may be used to derive the optical Bloch equations by adding the driving term $\sim \cos(\omega t)$ \textit{a posteriori} under the rotating wave approximation (RWA) \cite{Cohen_Tannoudji_atomphoton}. We here fully account for counter-rotating terms, which lead to additional sidebands in the atomic structure.

As evident from Eq. \eqref{eq:1}, the coupling of the atoms to a classical driving field gives rise to a periodically time-dependent $N$-atom Hamiltonian that can be treated non-perturbatively using Floquet theory \cite{Floquet:1883,Shirley:1965}. The quasi-stationary $N$-atom Floquet states of $\mathcal{H}_\mathrm{S}(t)$, indexed by $\alpha$, can be written as $\ket{\psi_\alpha(t)}=\exp(-i\mu_\alpha t)\,\ket{\phi_\alpha(t)}$, where $\mu_\alpha$ denotes the quasienergy, and $\ket{\phi_\alpha(t)}=\ket{\phi_\alpha(t+T)}$ is periodic with the period $T$ of the Hamiltonian $\mathcal{H}_\mathrm{S}(t)$. We denote by $\ket{\pm}=\exp(-i\mu_{\pm}t)\ket{\phi_\pm(t)}$ the Floquet states of a single driven two-level atom and by $\mu_{\pm}$ the corresponding quasienergies. 
Given the periodicity of $\ket{\phi_\alpha(t)}$, every Floquet state can be written as a Fourier series, $\ket{\psi_\alpha(t)}=\exp(-i\mu_\alpha t)\sum_{n\in\mathbb{Z}} \ket{\phi^{(n)}_\alpha}\exp(i n \omega t)$, where the index $n$ represents the number of photons emitted into or absorbed from the driving field \cite{Shirley:1965}. Each Floquet state thus comprises a ladder of levels (=sidebands) separated by $\omega$, with the Fourier amplitudes $\ket{\phi_\alpha^{(n)}}$ determining their population.

\emph{Master Equation. ---}
We note that the atom-reservoir interaction Hamiltonian $\mathcal{H}_\mathrm{I}$ is represented in the form $\mathcal{H}_\mathrm{I}=\sum_i A_i\otimes B_i$, with the system operators $A_i\equiv \sigma^x_i$ and the reservoir operators $B_i \equiv -i\sum_{\vec{k},s} g_{\vec{k}s}^{i} ( a_{\vec{k}s}-a^\dagger_{\vec{k}s})$. The periodic time dependence of $\mathcal{H}_\mathrm{S}$ necessitates a representation of $A_i$ in Floquet states \cite{Bluemel:1991,Mori_2023}, rather than in the eigenstates of $\mathcal{H}_\mathrm{S}$, as used for time-independent Hamiltonians. 
Furthermore, rapidly decaying bath correlation functions are required, which allows the application of the Born-Markov approximation \cite{Breuer:2007}. These assumptions are typically well justified in quantum optical systems. Casting the master equation in Lindblad form additionally requires the \emph{secular approximation}, for which the following hierarchy must be satisfied: $1/\omega,\,\tau_\mu\ll \tau_{cg}\ll \tau_s$, with $\tau_\mu$ a time scale pertaining to the inverse of differences of quasienergy spacings \cite{suppl}, $\tau_{cg}$ denoting the master equation's coarse-graining time \cite{Cohen_Tannoudji_atomphoton} and $\tau_s$ the characteristic time scale on which the system's state changes appreciably (in the interaction picture w.r.t. $\mathcal{H}_\mathrm{S}+\mathcal{H}_\mathrm{R}$) under the influence of the reservoir coupling. Physically, this means that the reservoir actually induces couplings between Floquet levels, on time scales which are longer than those of the intrinsic (unitary) Floquet dynamics.
Let us also note that (quasi)degeneracies of quasienergies can lead to the invalidity of the secular approximation \cite{suppl}. Henceforth, we assume that the system's quasienergies are arranged s.t. $\tau_\mu\ll \tau_{cg}\ll\tau_s$.

The derivation of the master equation requires evaluation of the spectral
correlation tensor of the reservoir operators associated with atoms $i$ and $j$
\cite{suppl, Breuer:2007},
\begin{align}
	R_{ij}(\nu) & =\int_{0}^{\infty}{\rm d}\tau \,{\rm Tr}_{\rm R}\{B_{i}(\tau)B_{j}(0)\rho_{\rm R}\}e^{i\nu \tau}\nonumber \\
	            & =\frac{1}{2}\Gamma_{ij}(\nu)+i\Omega_{ij}(\nu), \label{tensor_phi}
\end{align}
with $\rho_{\rm R}=\exp(-\beta \mathcal{H}_{\mathrm{R}})/\mathrm{Tr}\{\exp(-\beta
\mathcal{H}_{\mathrm{R}})\}$ the density operator of the reservoir ($\beta\equiv1
/k_{\rm B}T_{\rm R}$), $\Gamma_{ij}(\nu)$ the atomic spontaneous ($i=j$) and
collective ($i\neq j$) decay rates, and $\Omega_{ij}(\nu)$ the dipole-dipole interactions
($i\neq j$) and Lamb shifts ($i=j$). The functions $\Gamma_{ij}(\nu)$ and
$\Omega_{ij}(\nu)$ yield standard expressions that coincide with those derived
in Ref. \cite{Lehmberg:1970}. In particular, $\Tilde{\Omega }_{ij}(\omega)=\Omega_{ij}
(\omega)+\Omega_{ij}(-\omega)$ corresponds to the classical interaction energy between
two dipoles $i$ and $j$.

In its general form, the resulting Floquet-Markov Lindblad master equation reads
\begin{widetext}
	\begin{equation}
		\dot{\rho}=-i[\mathcal{H_\mathrm{dp}},\rho]+\sum_{i,j}\sum_{m\in\mathbb{Z}}\sum
		_{\{\Delta \mu\}}\Gamma_{ij}(m\omega +\Delta\mu)\Big[D_{m}^{i}[\Delta\mu]\rho
		D_{m}^{j}[\Delta\mu]^{\dagger}-\frac{1}{2}\{\rho,D_{m}^{j}[\Delta\mu]^{\dagger}
		D_{m}^{i}[\Delta\mu]\}\Big],\label{meq}
	\end{equation}
\end{widetext}
where
\begin{equation}
	\mathcal{H}_{\mathrm{dp}}=\sum_{i,j}\sum_{m\in\mathbb{Z}}\sum_{\{\Delta \mu\}}\Omega
	_{ij}(m\omega +\Delta\mu)D_{m}^{j}[\Delta\mu]^{\dagger} D_{m}^{i}[\Delta\mu],\label{Hdp}
\end{equation}
with
$D_{m}^{i}[\Delta\mu]=\sum_{\mu_\beta-\mu_\alpha=\Delta\mu}\langle\!\langle \phi_{\alpha}
|A_{i}|\phi_{\beta}\rangle \!\rangle_{m}\ket{\psi_\alpha}\bra{\psi_\beta}$, $\langle
\!\langle \phi_{\alpha}|A_{i}|\phi_{\beta}\rangle \!\rangle_{m}=\int_{0}^{T}{\rm d}
t \braket{\phi_\alpha|A_i|\phi_\beta}e^{i m \omega t}/T$, and $\{\Delta\mu\}$
the set of all pairwise differences of quasienergies. \textcolor{black}{While
the master equation exhibits no explicit time dependence in the interaction
picture, the operators $D_{m}^{i}[\Delta\mu]$ inherit the Floquet states' time
dependence in the Schrödinger picture, which, as remarked in Ref. \cite{Breuer:2007},
is an intuitive consequence of the (periodic) distortion of the atomic dipole
moments due to the strong field \cite{Buchleitner_1994}, and the atoms' coupling to
the reservoir through their dipole moment.}

We next discard the Lamb shifts to focus on the Hamiltonian of dipolar interactions ${\cal H}_{\rm dp}$. Restricting ourselves to the two-atom case, $N=2$, and dropping the atomic indices $i,j$ for brevity, the atom-atom dipole interaction Hamiltonian reads
\begin{align}
    \mathcal{H}_\mathrm{dp}^{(2)}&=c_{++}\left(\ket{++}\!\bra{++}-\ket{+-}\!\bra{+-}\right.\nonumber\\
    &\left.\,\quad\quad-\ket{-+}\!\bra{-+}+\ket{--}\!\bra{--}\right) \nonumber\\
    &\,+c_{+-}\left(\ket{+-}\!\bra{-+}+\ket{-+}\!\bra{+-}\right),\label{eq:6} 
\end{align}
with coefficients
\begin{align}
c_{++}&=\sum_{m\in\mathbb{Z}}|\langle\!\langle\phi_+|\sigma^x|\phi_+\rangle\!\rangle_m|^2 \,\Tilde{\Omega}(m\omega),\label{eq:7}\\
    c_{+-}&=\sum_{m\in\mathbb{Z}}  |\langle\!\langle \phi_-|\sigma^x|\phi_+\rangle\!\rangle_m|^2\,\Tilde{\Omega}(\mu_+-\mu_- +m\omega).\label{eq:8}
\end{align}

\noindent Strong driving results in non-vanishing sideband amplitudes, energetically separated by $m\omega$, of the Floquet states, allowing for the resonant absorption and emission of the excitations from and into the reservoir at multiple discrete energies rather than only at the bare atom's transition frequency $\omega_{eg}$. This is formally reflected by the increasing number of relevant energies in the sum in Eqs. \eqref{eq:7}-\eqref{eq:8} or in the decay rates (Eq. \eqref{eq:decayrates} in \cite{suppl}).

\begin{figure}[b]
\includegraphics[width=\linewidth]{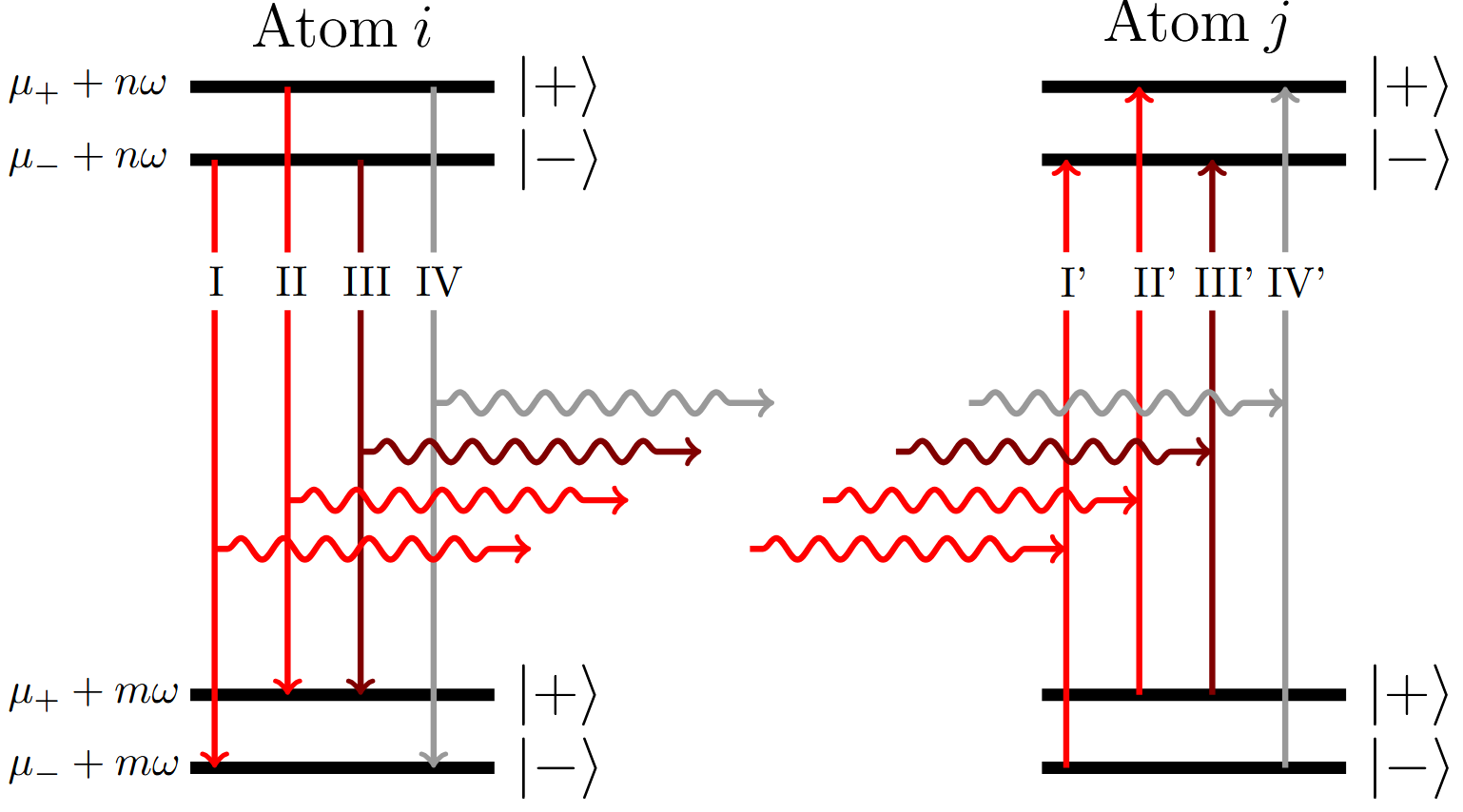}
\caption{\label{fig:couplings} Two steps on the Floquet ladder separated by $(n-m)\omega$, revealing possible resonant couplings between the atoms. The resonant interactions between the transitions ${\rm I}-{\rm I}^\prime$, ${\rm I}-{\rm II}^\prime$ and ${\rm II}-{\rm II}^\prime$, ${\rm II}-{\rm I}^\prime$ (red) correspond to the first and second line in Eq. \eqref{eq:6}, with coupling strength (8), whereas the resonant interactions between transitions ${\rm III}-{\rm III}^\prime$ (burgundy) and ${\rm IV}-{\rm IV}^\prime $ (gray) give rise to the third line in Eq. \eqref{eq:6}, with strength (9).}
\end{figure}

The physical meaning of the Hamiltonian ${\cal H}_{\rm dp}^{(2)}$ is illustrated in Fig.~\ref{fig:couplings}, which depicts a subset of the Floquet spectra of two interacting atoms. Associating the two-atom Floquet states $\ket{++}$, $\ket{+-}$, $\ket{-+}$, and $\ket{--}$ in Fig.~\ref{fig:couplings} with the terms of the Hamiltonian \eqref{eq:6}, we observe that the latter describe resonant couplings between the Floquet states. Transitions I and II of atom $i$ can each couple to two transitions (I' and II') of atom $j$. Likewise, transitions III and IV of atom $i$ can each couple to precisely one transition (III' or IV') of atom $j$. Note that the resonance character of the couplings of $H_{\rm dp}^{(2)}$ is an immediate consequence of the secular approximation, and that the underlying single-atom structure depicted in Fig. \ref{fig:couplings} is, in the limit of weak driving, the same as the one giving rise to the Mollow triplet in single atom resonance fluorescence \cite{Mollow_69}.

\emph{Simulation of spin models. ---}
    The dipole-dipole interaction Hamiltonian \eqref{eq:6} opens new avenues for probing hitherto unexplored dynamical regimes of strongly driven, interacting many-body systems, on account of the many sidebands of the Floquet states $\ket{\pm}$. Since the latter can in general be evaluated only numerically, we will illustrate a corollary of Eq.~\eqref{eq:6} for weak driving. Here, under the RWA,  the Floquet states are analytically well approximated  \cite{Meystre_2007} by the semiclassical dressed states $\ket{+}^{sc}=\exp(-i\mu_+t)(\sin(\theta/2)e^{i\omega t}\ket{g}+\cos(\theta/2)\ket{e})$ and $\ket{-}^{sc}=\exp(-i\mu_-t)(\cos(\theta/2)e^{i\omega t}\ket{g}-\sin(\theta/2)\ket{e})$, with $\cos(\theta)=-\delta/\sqrt{\delta^2+\Omega_\mathrm{R}^2}$, $\mu_\pm=(\omega \pm \sqrt{\Omega_{\rm R}^2 +\delta^2})/2$ and $\delta=\omega-\omega_{eg}$, where the time dependence can be removed by a simple unitary transformation. Within the validity regimes of the RWA and of Eq. \eqref{eq:6} (also see Fig. \ref{fig:4a} in \cite{suppl}), we can substitute $\ket{\pm}^{sc}$ in the $N$-atom generalization of Eq. \eqref{eq:6}, to obtain
    \begin{align}
        \mathcal{H}_\mathrm{dp}&=\sum_{i\neq j}\sum_{\alpha,\beta\in\{x,y,z\}}\, \sigma_i^\alpha J_{\alpha\beta}^{ij}\, \sigma_j^\beta, \label{eq:spin}
    \end{align}
    with $\sigma^\alpha_i$ the standard Pauli matrix of atom $i$ in the bare atomic basis, and $J_{\alpha\beta}$ the interaction strengths (Eqs. \eqref{eq:jxx}-\eqref{eq:jxz} in \cite{suppl}) that can be controlled by changing the Rabi frequency $\Omega_{\rm R}$ and detuning $\delta$.
    Hamiltonian \eqref{eq:spin} corresponds to the anisotropic Heisenberg model, which plays
    a prominent role in simulations of many-body spin systems with Rydberg atoms \cite{Browaeys_2020}.
    While the realization of such interactions within the paradigm of Floquet engineering typically involves sophisticated protocols \cite{Weitenberg:2021,Goldman:2015,Geier:2021,Scholl:2022}, a tunable variety of spin models becomes accessible simply through  periodic driving of the atoms. Note that Hamiltonian \eqref{eq:spin} can also be obtained through temporal coarse-graining of the standard flip-flop operator of dipole-dipole interactions, $\propto \ket{eg}\!\bra{ge}+\ket{ge}\!\bra{eg}=\sigma^x_i\otimes\sigma^x_j+\sigma^y_i\otimes\sigma^y_j$ \cite{suppl}.
    This again suggests that the full potential of the dipolar interactions derived in our present contribution can unfold in the strong driving limit.

To conclude, let us note that the required separation of time scales and the validity regime of  Eqs. \eqref{eq:6} and  \eqref{eq:spin} are easily attainable with Rydberg states. For typical parameters $\omega_{eg}\sim 10\, \si{\giga\hertz}$, $d_{eg}\sim 1000\, ea_0$ \cite{Gallagher_1994}, and for the experimentally feasible values $r_{ij}\sim 40\, \si{\micro\meter}$, $\Omega_{\rm R}\sim 0.01\omega_{eg}$ \cite{Lee:2017,Ditzhuijzen:2008} and $\delta=0$, the above hierarchy of time scales is established, and a dipole-dipole interaction strength 
$\tilde{\Omega}_{ij}\sim 96\,\si{\kilo\hertz}$ implements the Heisenberg model with $J_{xx}=2J_{yy}=2J_{zz}=48 \,\si{\kilo \hertz}$. 

\vfil

\bibliography{lit_DIF}

\clearpage
\section{Supplemental Material}
\subsection{Decay rates, jump operators and dipole-dipole interaction energy}
\label{supp:decayrates}
\noindent
For a reservoir at temperature $T_{\rm R}$, the collective decay rate is given by
\begin{align}
	\Gamma_{ij}(\omega)=\Gamma_{ij}'(|\omega|)\Big[\Theta(\omega)(1+n(|\omega|))+ \Theta(-\omega)n(|\omega|)\Big],
\end{align}
where $n(\omega)=1/[\exp(\omega/(k_{\rm B}T_{\rm R}))-1]$ is the thermal occupation
and $\Theta(\omega)$ denotes the Heaviside function. Further, we have defined an
auxiliary single atom spontaneous decay rate $(i=j)$ as
\begin{align}
	\Gamma_{ii}' & =\frac{\mu^{2}\omega^{3}}{3 \pi \epsilon_{0} c^{3} },
\end{align}
and an auxiliary collective decay rate $(i\neq j)$
\begin{align}
	\Gamma_{ij}'(\omega) & =\frac{\mu^{2}\omega^{3}}{2 \pi \epsilon_{0} c^{3} }\Bigg[\left(1-\cos^{2}(\theta)\right)\frac{\sin(\xi)}{ \xi}\nonumber \\
	                     & +\left(1-3\cos^{2}(\theta)\right)\left(\frac{\cos(\xi)}{\xi^{2}}-\frac{\sin(\xi)}{\xi^{3}}\right)\Bigg],
\end{align}
with $\xi:=\omega r_{ij}/c$, the interatomic distance $r_{ij}$ and the angle
$\theta$ between the dipole moments and the interatomic axis. The dipole-dipole interaction
energy reads
\begin{align}
	\Tilde{\Omega}_{ij}(\omega) & = \frac{\mu^{2} \omega^{3} }{4\pi\epsilon_{0} c^{3}}\Bigg[ -\left(1-\cos^{2}(\theta)\right) \frac{\cos(\xi)}{\xi}\nonumber \\
	                            & + \left(1-3\cos^{2}(\theta)\right) \left(\frac{\sin(\xi)}{\xi^{2}}+\frac{\cos(\xi)}{\xi^{3}}\right)\Bigg]\,.
\end{align}
These are familiar expressions previously derived in the study of systems of unperturbed
atoms, see Ref. \cite{Lehmberg:1970}. We remark that the dipole-dipole interaction
contains both near- and far field terms, scaling as $1/r_{ij}^{2}$,
$1/r_{ij}^{3}$ and $1/r_{ij}$, respectively. For a pair of two-level atoms, the diagonal
form of the Lindblad master equation
\begin{align}
	\dot{\rho}=-i[\mathcal{H}_{\mathrm{dp}}^{(2)},\rho]+\sum_{k=1}^{6}\gamma_{k}\Big[L_{k} \rho L_{k}^{\dagger}-\frac{1}{2}\{ L_{k}^{\dagger} L_{k},\rho\}\Big]
\end{align}
can be readily written down and features the following decay rates and jump operators:
\vspace{3cm}
\begin{widetext}
	\begin{align}
		\label{eq:decayrates}\gamma_{1} & =\sum_{m\in \mathbb{Z}}|\langle\!\langle\phi_ +|\sigma^{x}|\phi_+\rangle\!\rangle_{m} |^{2} (\Gamma_{11}(m\omega)+\Gamma_{12}(m\omega)),\nonumber                         & L_{1} & =\sqrt{2}(\ket{++}\!\bra{++}-\ket{--}\!\bra{--}),\nonumber                      \\
		\gamma_{2}                      & =\sum_{m\in \mathbb{Z}}|\langle\!\langle\phi_ +|\sigma^{x}|\phi_+\rangle\!\rangle_{m} |^{2} (\Gamma_{11}(m\omega)-\Gamma_{12}(m\omega)),\nonumber                         & L_{2} & =\sqrt{2}(\ket{+-}\!\bra{+-}-\ket{-+}\!\bra{-+}),\nonumber                      \\
		\gamma_{3}                      & =\sum_{m\in \mathbb{Z}}|\langle\!\langle \phi_-|\sigma^{x}|\phi_+\rangle\!\rangle_{m} |^{2} (\Gamma_{11}(m\omega+\mu_{+}-\mu_{-}) +\Gamma_{12}(m\omega+\mu_{+}-\mu_{-})), & L_{3} & =\ket{-}\!\bra{+}\otimes \mathds{1}+\mathds{1}\otimes\ket{-}\!\bra{+},\nonumber \\
		\gamma_{4}                      & =\sum_{m\in \mathbb{Z}}|\langle\!\langle \phi_-|\sigma^{x}|\phi_+\rangle\!\rangle_{m} |^{2} (\Gamma_{11}(m\omega+\mu_{+}-\mu_{-})-\Gamma_{12}(m\omega+\mu_{+}-\mu_{-})),  & L_{4} & =\ket{-}\!\bra{+}\otimes\mathds{1}-\mathds{1}\otimes\ket{-}\!\bra{+},\nonumber  \\
		\gamma_{5}                      & =\sum_{m\in \mathbb{Z}}|\langle\!\langle \phi_+|\sigma^{x}|\phi_-\rangle\!\rangle_{m} |^{2} (\Gamma_{11}(m\omega-\mu_{+}+\mu_{-})+\Gamma_{12}(m\omega-\mu_{+}+\mu_{-})),  & L_{5} & =\ket{+}\!\bra{-}\otimes\mathds{1}+\mathds{1}\otimes\ket{+}\!\bra{-},\nonumber  \\
		\gamma_{6}                      & =\sum_{m\in \mathbb{Z}}|\langle\!\langle \phi_+|\sigma^{x}|\phi_-\rangle\!\rangle_{m} |^{2} (\Gamma_{11}(m\omega-\mu_{+}+\mu_{-}) -\Gamma_{12}(m\omega-\mu_{+}+\mu_{-})), & L_{6} & =\ket{+}\!\bra{-}\otimes\mathds{1}-\mathds{1}\otimes\ket{+}\!\bra{-}.
	\end{align}
\end{widetext}
Note that we used the symmetries $\Gamma_{11}(...)=\Gamma_{22}(...)$ and
$\Gamma_{12}(...)=\Gamma_{21}(...)$.
\subsection{Coupling constants in spin Hamiltonian}
\label{appendix:b} The matrix $\textbf{J}^{ij}$ in Eq. \eqref{eq:spin} has five nonzero
components, given by the elements
\begin{align}
	\label{eq:jxx}J_{xx}^{ij} & =\frac{1}{32}\tilde{\Omega}_{ij}(\omega) (3 \cos (4 \theta_{m} )+13),
\end{align}

and further
\begin{align}
	J_{yy}^{ij} & =\frac{1}{8}\tilde{\Omega}_{ij}(\omega) (\cos (2 \theta_{m})+3),                                                \\
	J_{zz}^{ij} & = \frac{1}{8}\tilde{\Omega}_{ij}(\omega) \sin^{2}(\theta_{m} ) (3 \cos (2 \theta_{m} )+5),                      \\
	J_{xz}^{ij} & =J_{zx}^{ij}= \frac{1}{32}\tilde{\Omega}_{ij}(\omega) (2 \sin (2 \theta_{m} )+3 \sin (4\theta_{m} )), \label{eq:jxz}
\end{align}
with dressed state mixing angle $\theta_{m}$ and driving frequency $\omega$. The
expressions were obtained using the approximation
$\tilde{\Omega}_{ij}(\omega + \mu_{+}-\mu_{-})\approx \tilde{\Omega}_{ij}(\omega)$, which holds for weak, near-resonant driving.
\vspace{.1cm}
\onecolumngrid

\begin{figure*}[t]
	\centering
    \subfigure[]{ \includegraphics[width=0.45\textwidth]{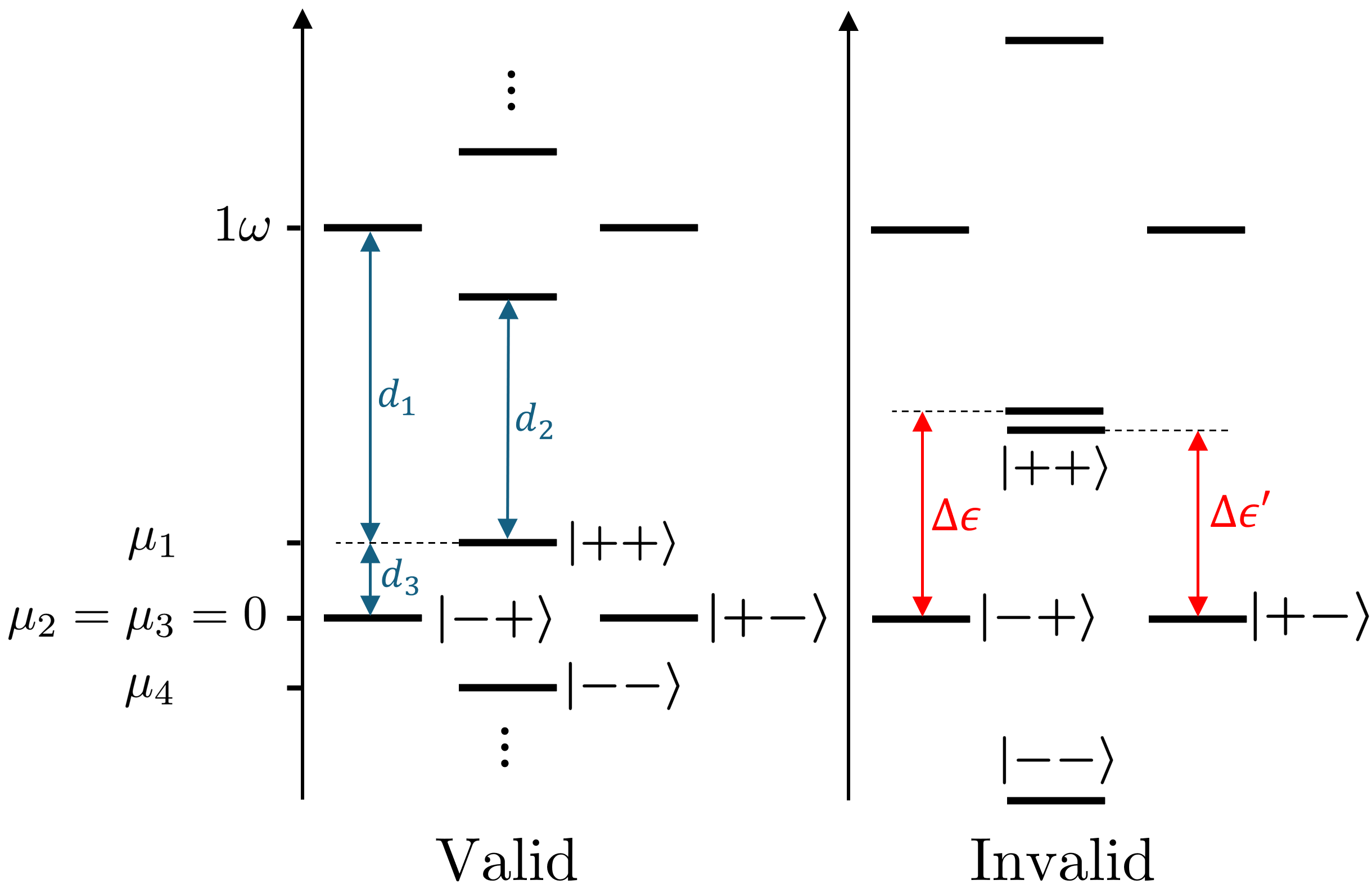} \label{fig:3a} }
	\hfill \subfigure[]{ \raisebox{7mm}{\includegraphics[width=0.45\textwidth]{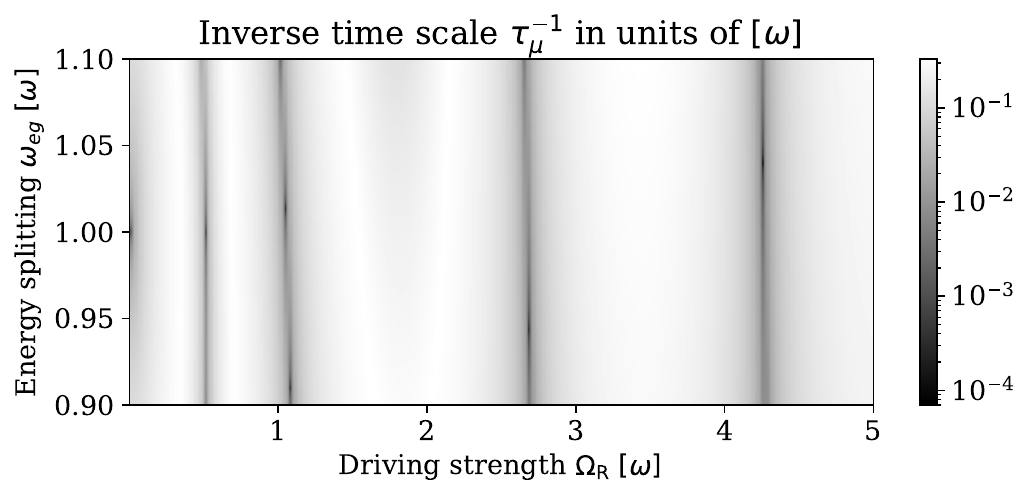}} \label{fig:3b} }
	\caption{(a) Two adjacent steps on the Floquet ladder, depicting the four two-atom
	Floquet states with corresponding two-atom quasienergies $\mu_{1}=2\mu_{+}$,
	$\mu_{2}=\mu_{3}=0$, $\mu_{4}=-\mu_{1}$. On the left-hand side, the levels are
	sufficiently separated; the distances pertaining to the inverse time scale
	$\tau_{\mu}^{-1}=\min(d_{1},d_{2},d_{3})$ are indicated. On the right-hand side,
	a quasidegeneracy leading to $\Delta\epsilon-\Delta\epsilon'\approx0$ for some
	$\Delta\epsilon, \,\Delta\epsilon'$ is illustrated qualitatively. (b) Illustration
	of regimes for which discarding of oscillating terms $\exp(i(\Delta\epsilon-\Delta
	\epsilon') t)$ with $\Delta\epsilon\neq \Delta\epsilon'$ is justified. Equation \eqref{eq:6}
	holds everywhere except in the dark shaded regions (where $\tau_\mu$ diverges), provided that the remaining separations of time scales, as outlined in the main text, persist.}
\end{figure*}
\twocolumngrid
\subsection{Secular approximation}
\label{appendix:C} The secular approximation consists in ignoring rapidly
oscillating terms $\propto \exp(i(\Delta\epsilon-\Delta\epsilon')t)$ in the
derivation of the master equation, with $\Delta\epsilon=\mu_{\alpha}-\mu_{\beta}+
m\omega$ and $\Delta\epsilon'=\mu_{\alpha'}-\mu_{\beta'}+m'\omega$. Here, we identify
regimes where this approximation is justified. A study of the Floquet spectrum's
relevant transition frequencies $\Delta\epsilon$ shows that this approximation
is valid for
\begin{gather}
	\tau_{\mu} \ll \tau_{cg}\ll \tau_{s},\nonumber\\ \quad \tau_{\mu}^{-1}=\min(|1-2|\mu_{+}||,|
	2\mu_{+}|,|1-4|\mu_{+}||), \label{eq:taumum}
\end{gather}
with $\mu_{+}$ the quasienergy of the single-atom Floquet state $\ket{+}$ (chosen in a Floquet zone
such that $2|\mu_{+}|<\omega$), $\tau_{cg}$ the master equation's coarse-graining time (that is, the smallest resolvable times scale of the master equation), and $\tau_{s}$ the characteristic time scale on which
the system evolves in the interaction picture w.r.t. $\mathcal{H}_{\rm R}+\mathcal{H}
_{\rm S}$. Note that Eq. \eqref{eq:taumum} holds irrespective of the number of
atoms. For a pair of atoms, a subset of the system's Floquet spectrum is
depicted in Fig. \ref{fig:3a}. Here, the quantity $\tau_{\mu}^{-1}$ describes the
minimal distance of quasienergies (disregarding the degeneracy of
$\mu_{2}=\mu_{3}$ stemming from the atoms' exchange symmetry, which was accounted for in the derivation of the master equation), i.e., $\tau_\mu^{-1}=\min(d_1,d_2,d_3)$, with $d_1,..,d_3$ defined in Fig. \ref{fig:3a}. In this case, the secular approximation is thus valid if the quasienergies are sufficiently separated.

\begin{figure*}[t]
	\centering
	\subfigure[]{ \includegraphics[width=0.45\textwidth]{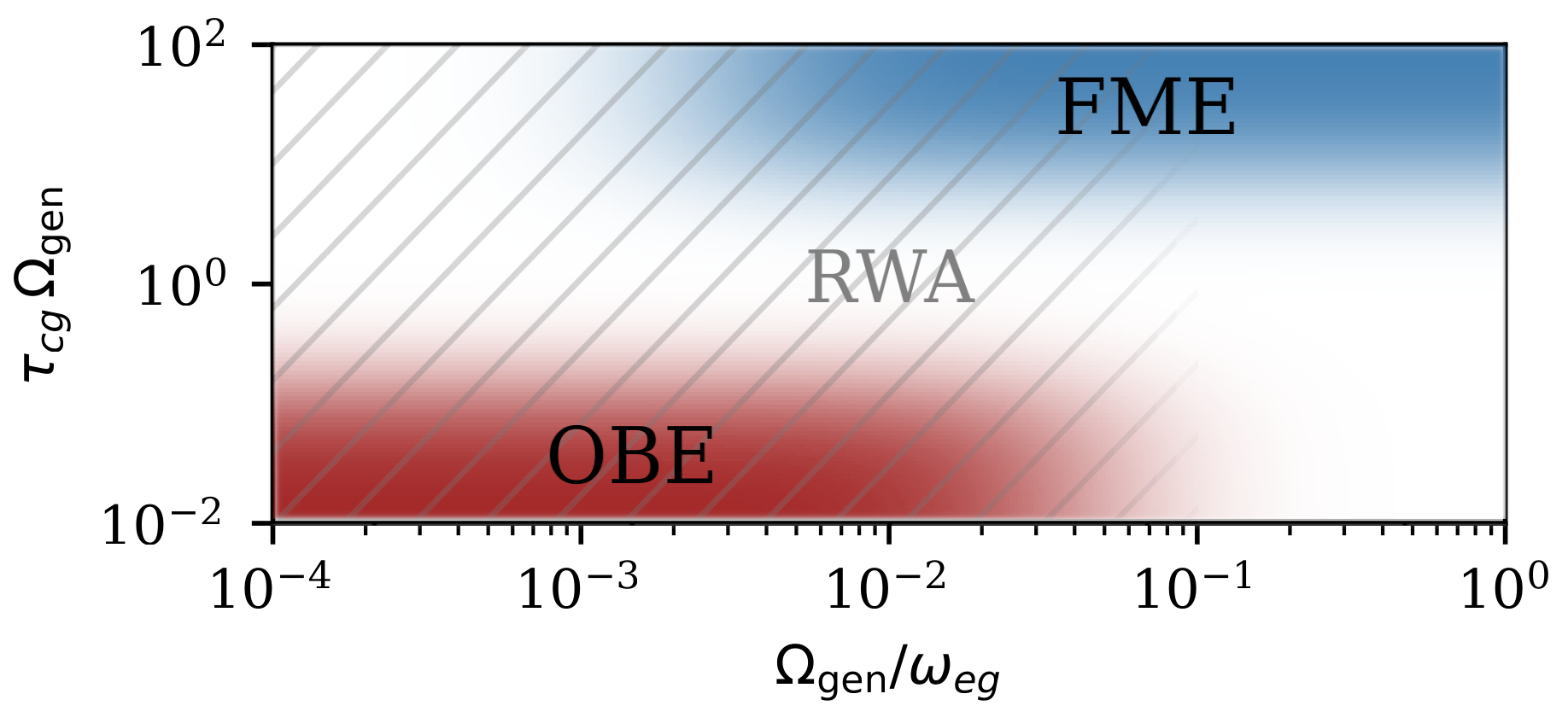} \label{fig:4a} }
	\hfill \subfigure[]{ \raisebox{11mm}{\includegraphics[width=0.45\textwidth]{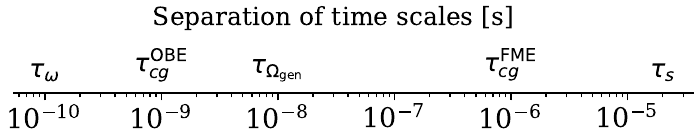} }\label{fig:4b} }
	\caption{(a) Qualitative illustration of regimes of validity of OBE and FME in the RWA regime (hatched lines)   (b) Separation of time scales for the exemplary values $\omega_{eg}\sim 10\, \si{\giga\hertz}$, $d_{eg}\sim 1000\, ea_0$, $r_{ij}\sim 40\, \si{\micro\meter}$, $\Omega_{\rm R}\sim 0.01\omega_{eg}$, $\delta=0$, $\vec{d}_{eg} \perp\vec{r}_{12}$, and $T_{\rm R}=0$ in the main text. We have $ \tau_\omega=1/\omega$, $\tau_\mu \sim 1/\Omega_{\rm gen}=\tau_{\Omega_{\rm gen}}$, and $\tau_s\sim 1/\tilde{\Omega}_{ij}(\omega_{eg})$. As remarked, the separation allows one to choose $\tau_{cg}\ll 1/\Omega_{\rm gen}$ or $\tau_{cg}\gg 1/\Omega_{\rm gen}$, giving rise to the $N$-atom OBE and FME with distinct expressions for the dipolar interaction. However, this is fully consistent, as another coarse-graining of the flip-flop interaction results in the structure presented in Eq. \eqref{eq:spin}.}
\end{figure*}

The inverse time scale $\tau^{-1}_{\mu}$ was computed for a parameter space spanned
by the driving strength and the atom's energy splitting in Fig. \ref{fig:3b}. Along
black stripes, $\tau_{\mu}$ diverges. Note in particular the violation of the inequality in Eq.
\eqref{eq:taumum} as $\Omega_{\rm R}$ tends to zero. This condition follows immediately from $\tau_\mu^{-1}\sim \Omega_{\rm gen}=\sqrt{\delta^2 + \Omega_{\rm R}^2}$ for $\Omega_{\rm gen} \ll \omega$. Consequently, the Floquet-Markov master equation (FME) does not remain valid for arbitrarily weak driving (see also Fig. \ref{fig:4a}).

\subsection{OBE vs. FME in RWA regime}

For time scales hierarchy $1/\omega\ll1/\Omega_{\rm gen}\ll\tau_s$ (RWA regime), the $N$-atom optical Bloch equations (OBE) and the flip-flop operator are known to provide an accurate description of the system \cite{Binninger:2019}, raising the question of how the flip-flop interaction can be reconciled with the seemingly more complicated interaction of Eq. \eqref{eq:spin}. The key to the answer is the coarse-graining time $\tau_{cg}$ used in the derivation of the master equation. With the above separation of time scales, two choices can be made, $\tau_{cg}\ll 1/\Omega_{\rm gen}$ or $\tau_{cg}\gg 1/\Omega_{\rm gen}$, giving rise to the OBE or FME, respectively \cite{Elouard_2020}. 
This implies that Eq. \eqref{eq:spin} can also be obtained by averaging the dynamics induced by the flip-flop interaction, as can be quickly verified. Consider two atoms with the flip-flop interaction 
\begin{align}
    \mathcal{H}^{(2)}_{\rm dp}=\tilde{\Omega}_{12}(\omega_{eg}) (\sigma^+\otimes\sigma^-+\sigma^-\otimes\sigma^+),
\end{align}
with the raising and lowering operators $\sigma^+=\ket{e}\!\bra{g}$ and $\sigma^-=\ket{g}\!\bra{e}$, respectively.
Moving to the interaction picture w.r.t. $\mathcal{H}_\mathrm{S}^{\rm RWA}+\mathcal{H}_\mathrm{R}$, where $\mathcal{H}_\mathrm{S}^{\rm RWA}=\frac{1}2{}\sum_i[\Omega_{\rm R}(\sigma_i^+e^{-i\omega t}+\sigma_i^-e^{i\omega t})+ \omega_{eg}\sigma_i^z]$, we can write \cite{Elouard_2020}
\begin{align}
    \sigma^{\pm}_I(t)=e^{\pm i\omega t}\sum_{\Delta\mu=0,\Omega_{\rm gen},-\Omega_{\rm gen}}\sigma^{\pm}( \Delta\mu)e^{ \pm i\Delta\mu t},
\end{align}
where
\begin{align}
    &\sigma^\pm(0)=\cos(\theta_m/2)\sin(\theta_m/2) \big(\ket{+}^{sc}\!\bra{+}^{sc}-\ket{-}^{sc}\!\bra{-}^{sc}\big),\nonumber\\
    &\sigma^\pm(\Omega_{\rm gen})=\cos^2(\theta_m/2) \ket{\pm}^{sc}\!\bra{\mp}^{sc},\nonumber\\
    &\sigma^\pm(-\Omega_{\rm gen})=-\sin^2(\theta_m/2) \ket{\mp}^{sc}\!\bra{\pm}^{sc},
\end{align}
with semiclassical dressed states $\ket{\pm}^{sc}$ and corresponding quasienergies $\mu_\pm=\omega/2\pm\Omega_{\rm gen}/2$ from the main text.
The coarse-grained dipole-dipole interaction Hamiltonian is obtained in the interaction picture through 
\begin{align}
    \mathcal{H}^{(2)}_{\rm dp-cg}(t)=\frac{1}{\tau_{cg}^{\rm FME}}\int_t^{t+\tau_{cg}^{\rm FME}}dt' \,\mathcal{H}_{\rm dp}^{(2)}(t').
\end{align}
If we now neglect phases in $\mathcal{H}_{\rm dp}^{(2)}(t')$ oscillating as $\pm\Omega_{\rm gen}$, $\pm 2\Omega_{\rm gen}$, on account of $1/\Omega_{\rm gen}\ll\tau_{cg}^{\rm FME}$, we obtain
\begin{align}
    \mathcal{H}^{(2)}_{\rm dp-cg}&=c_{++}^{cg}\left(\ket{++}^{sc}\!\bra{++}^{sc}-\ket{+-}^{sc}\!\bra{+-}^{sc}\right.\nonumber\\
    &\left.\,\quad\quad-\ket{-+}^{sc}\!\bra{-+}^{sc}+\ket{--}^{sc}\!\bra{--}^{sc}\right) \nonumber\\
    &\,+c_{+-}^{cg}\left(\ket{+-}^{sc}\!\bra{-+}^{sc}+\ket{-+}^{sc}\!\bra{+-}^{sc}\right)
    \label{Hcg},\end{align}
with $c_{++}^{cg}=2\tilde{\Omega}_{12}(\omega_{eg})\cos^2(\theta_m/2)\sin^2(\theta_m/2)$ and $c_{+-}^{cg}=\tilde{\Omega}_{12}(\omega_{eg})(\sin^4(\theta_m/2)+\cos^4(\theta_m/2))$, which is precisely what is immediately obtained from Eq. \eqref{eq:6} upon substitution of the dressed states and exploiting the sufficient flatness of the spectral correlation tensor around $\omega_{eg}$, such that $\tilde{\Omega}_{ij}(\omega_{eg}\pm \Omega_{\rm gen})\approx \tilde{\Omega}_{ij}(\omega_{eg})$.

\onecolumngrid
\twocolumngrid

\end{document}